\documentclass{elsart}
\usepackage{graphicx}
\begin{document}
\begin{frontmatter} 

\title{Statistical modelling of electrochemical deposition of nanostructured
hybrid films with ZnO-Eosin Y as a case example}
\author[uff]{F. D. A. Aar\~ao Reis\corauthref{cor1}}
\ead{reis@if.uff.br}
\corauth[cor1]{Fax number: (55) 21-2629-5887 }
\author[pmc]{J. P. Badiali}
\ead{badiali@ccr.jussieu.fr}
\author[enscp]{Th. Pauport\'e}
\ead{thierry-pauporte@enscp.fr}
\author[enscp]{D. Lincot}
\ead{daniel-lincot@enscp.fr}
\address[uff]{
Instituto de F\'\i sica, Universidade Federal Fluminense,
Avenida Litor\^anea s/n, 24210-340 Niter\'oi RJ, Brazil}
\address[pmc]
{Laboratoire d'Electrochimie et de Chimie Analytique (UMR 7575 - CNRS - ENSCP -
Paris 6), Universit\'e P. et M. Curie, 4 Place Jussieu, 75005 Paris, France}
\address[enscp]
{ Laboratoire d'Electrochimie et de Chimie Analytique (UMR 7575 - CNRS - ENSCP -
Paris 6), \'Ecole Nationale Sup\'erieure de Chimie de Paris, 11 rue Pierre
et Marie Curie, 75231, Paris, France}
\date{\today}
\maketitle

\begin{abstract}
We study models
of electrodeposition of hybrid organic-inorganic films with
a special focus on the growth of ZnO with eosin Y.
First we propose a rate equation model which assumes that the organic
additives form branches with an exposed part above the $ZnO$ deposit, growing
with larger rate than the pure film and producing $ZnO$ at the exposed length.
This accounts for the generation of ${OH}^-$ ions from reduction of dissolved
oxygen near the branches and reactions with ${Zn}^{2+}$ ions to form $ZnO$
molecules. The film grows with the same
rate of the branches, which qualitatively explains their catalytic effect, and
we discuss the role of the additive concentration.
Subsequently, we propose a statistical model which represents
the diffusion of the hydroxide precursor and of
eosin in solution and adopt simple probabilistic rules for the reactions,
similarly to diffusion-limited aggregation models.
The catalytic effect is represented by the preferencial
production of ${OH}^-$ ions near eosin.
The model is simulated with relative concentrations in
solution near the experimental values. An improvement of the growth rate is
possible only with a rather large apparent diffusion coefficient of eosin in
solution compared to that of hydroxide precursors. When
neighboring eosin clusters competitively grow, the increase in the growth
rate and a high eosin loading are observed in the simulated deposits. Those
features are in qualitative agreement with experimental results. 
\end{abstract}


\end{frontmatter}

\section{Introduction}

The growth of thin films from solutions has emerged as an efficient, low
temperature, versatile preparation route which can be used at large scale for
the production of high quality materials \cite{revlincot}. One of the interests
of these methods
is the possibility of adding foreign soluble compounds to the deposition bath
and modulating film properties by playing with the interactions which arise
between the growing film and these compounds. Among the additives, organic ones
are of utmost interest since many of them have been shown to act as templating
agents and/or crystal growth directing agents, and in many cases as
functionalizing agents for the deposits. 

These effects are well-documented in the case of zinc oxide. This inorganic
compound has attracted much attention due to a broad range of potential high
technology applications such as surface acoustic wave filter \cite{emanetoglu},
light emitting diodes \cite{konenkamp}, lasers \cite{huang}, varistors
\cite{lin}, gas sensors \cite{golego} and solar cells \cite{keis}. $ZnO$ can be
prepared as a thin film by chemical deposition methods \cite{vaysieres,tian} or
by electrodeposition
\cite{izaki,peulon1,peulon2,pauporte1,pauporte2,pauporte3}. The dramatic
effects of organic additives on film structures and morphologies have been
reported with both deposition routes. For instance, in the presence of citrate,
chemically deposited $ZnO$ are formed of stacked nanoplates assembled in a
biomimetic manner \cite{tian}. However the most impressive morphological
changes have been observed in electrodeposited zinc oxide. In the presence of
sodium dodecylsulfate (SDS), the formation of lamellar nanostructures has been
shown \cite{choi,michaelis}. The SDS can be removed from the film and highly
porous thin films are released \cite{michaelis}. Different dyes, such as eosin
Y (EY) \cite{yoshida1,pauporte4,yoshida2,pauporte5}, fluorescein (FL)
\cite{okabe}, Tetrasulfonated phthalocyanines (TSPc)
\cite{yoshida3,yoshida4,schlettwein,pauporte6} and riboflavin (RI)
\cite{karuppuchamy}, have also been shown to incorporate in the film and to
give rise to nanostructures. These organic dyes contain negatively charged
functions (carboxylate (EY, FL), sulfonate (TSPc) or phosphonate (RI) groups)
which allow their direct binding to the oxide crystal surface during the
synthesis.

While the effects of additives in film shaping are widely illustrated
in experimental works, particularly that of the organic ones, the exact role of
these compounds in the growth process remains to be clarified in most cases.
The aim of the present work is to fill this gap by
proposing kinetic and statistical models for zinc oxide electrodeposition in
the presence of EY in solution, which focus on a small number of basic
features of those processes. The starting point for this modeling is a series of
experimental results, some of them also observed in deposition with different
organic additives \cite{michaelis}. 

The first step of electrochemical zinc oxide synthesis is the cathodic
reduction of a hydroxide precursor such as molecular oxygen
\cite{peulon1,pauporte1}, hydrogen peroxide \cite{pauporte2,pauporte3} or
nitrate ions \cite{izaki,yoshida1}:
\begin{equation}
\frac{1}{2} O_2 + H_2O + 2e^- \rightarrow  2{OH}^-  ,
\label{reacoxigen}
\end{equation}
\begin{equation}
H_2O_2 + 2e^- \rightarrow 2{OH}^- 
\label{reacperoxide}
\end{equation}
or
\begin{equation}
{NO_3}^- + H_2O + 2e^- \rightarrow {NO_2}^- + 2{OH}^- .
\label{reacnitrate}
\end{equation}
At a temperature slightly above room temperature \cite{goux}, well crystallized
zinc oxide with the wurtzite hexagonal structure is precipitated at the
electrode surface:
\begin{equation}
Zn^{2+} + 2 {OH}^- \rightarrow ZnO + H_2O.
\label{reaczno}
\end{equation}
In many cases, it has also been shown that the organic component presents
catalytic properties for the deposition process. This catalytic effect is
connected to the improvement of $O_2$ reduction (Eq. \ref{reacoxigen}) with EY
\cite{yoshida2}, TSPc \cite{pauporte6}, SDS \cite{michaelis}, whereas EY has
also been shown to catalyze the $H_2O_2$ reduction (Eq. \ref{reacperoxide})
\cite{pauporte4}. Thus, one of the assumptions of our models is a distinguished
rate of formation of $ZnO$ near the species that represent aggregated eosin
molecules.

The zinc oxide films
prepared in the presence of EY show large round shaped single crystals of zinc
oxide filled with self-assembled dye aggregates, as shown in Fig. 1
\cite{pauporte5}. It is observed that the dye can be almost completely
removed by a soft chemical
treatment \cite{yoshida5,pauporte7}, revealing a network of
mesopores which is connected to the surface and can be filled with a
solution. This network, formed by aggregated dye molecules, acts as a template
for the growth of $ZnO$. In our models, these experimental facts justify the
assumption of formation of connected branches of the species that represent the
additive molecules in the films. 

In Sec. II of this paper, we will introduce our first model, which is based on
rate equations, and explains some
features of the cooperative growth of a $ZnO$ film and additive aggregates. This
model is particularly useful for understanding the role of additive
concentrations and diffusion coefficients in a qualitative way. Subsequently, in
Sec. III we will present a statistical model which
represents the main microscopic features of those processes by adopting
probabilistic
rules of diffusion and aggregation of some chemical species in solution and in
the deposit. The hypothesis of a growth mechanism controlled by
diffusion-limited aggregation of the eosin molecules was anticipated in Ref.
\protect\cite{lincot1}. Simulation results of that statistical model, presented
in Sec. IV, provide estimates
of growth rates, structures of $ZnO$ and EY deposits and shows a role of
diffusion coefficients which are in qualitative agreement with experimental
findings. Finally, in Sec. V we
summarize our results and present our conclusions.

\section{Rate equation model}

This model aims at explaining basic kinetic features of the growth of $ZnO$ with
organic additives with drastic simplifications of the film structure.

First we assume that pure
$ZnO$ growth takes place with rate $r_0$ (in nanometers per second), due to
reactions of type (\ref{reaczno}) near the film surface. If the average volume
occupied by one molecule is $V$, then $r_0/V$ is the growth rate per unit area
in that case. Even in the presence of the additives, the roughness of the $ZnO$
film surface will be neglected.

When the additive is present, we assume that it forms straight structures which
grow from the electrode, as shown in Fig. 2. This is consistent with the
experimental finding of porous structures when eosin Y was removed from the
hybrid films \cite{yoshida5,pauporte7}, but with an initial oversimplification
of that structure. The total number of linear
branches of the additive is $N_0$, which depends on the nucleation  of the
additive at the electrode surface in the beginning of the growth process. Each
branch is assumed to grow with rate $r_A$ if it is not covered by the $ZnO$
deposit ($r_A>r_0$), so that the height of each branch increases as $H=r_A t$.
$r_A$ is expected to increase with the concentration $\rho_A$ and the
diffusion coefficient $D_A$ of the additive in solution, but the exact
dependence on those quantities will not be important at this point.

In order to represent the catalytic effect of the additive, we assume that an
excess of ${OH}^-$ is produced near the branches, which leads to the
production of $ZnO$ at those regions. The rate of production of $ZnO$ is
expected to be proportional to the length of those branches which is exposed
above the film surface, $\left( H-h\right)$ (see Fig. 2), where $h$ is the
$ZnO$ film height. Consequently, the total rate of production of $ZnO$
molecules near the additive branches is $N_0k\left( H-h\right)$, where $k$ is a
reaction constant. A constant reaction rate in this model accounts for
the assumption that the oxygen reduction reaction is not under diffusion
control. Also, it is assumed that the $ZnO$ molecules immediately
precipitate and aggregate to the flat film surface, leaving the branches
uncovered.

Under these conditions, if the area of the film is $A$, then the number of $ZnO$
molecules deposited per unit time is
\begin{equation}
\frac{dN}{dt} = \frac{r_0}{V} A + N_0 k { \left( H-h\right)} .
\label{dNdt}
\end{equation}
Here, the first contribution in the right hand side comes from the reactions
ocurring near the film surface and the second one comes from the reactions near
the branches. The growth rate of the $ZnO$ deposit is
$\frac{dh}{dt} = \frac{V}{A} \frac{dN}{dt}$, which gives
\begin{equation}
\frac{dh}{dt} = r_0 -ah +r_A at \qquad ,\qquad a\equiv \frac{N_0kV}{A} .
\label{dhdt}
\end{equation}

The solution of Eq. (\ref{dhdt}) is
\begin{equation}
h = r_A t + \frac{\left( r_A-r_0\right)}{a} {\left( e^{-at}-1 \right)} .
\label{ht}
\end{equation}
It gives a constant growth rate $r_A$ for the deposit at long times, with  $1/a$
being the characteristic time of decay. It means that the film growth rate
attains the same growth rate of the branches in the steady state. The rate
of production of $ZnO$ near the film surface is $r_0<r_A$, but the exposed
height of the branches saturates at $H-h = \frac{r_A-r_0}{a}$, and produces an
excess of $ZnO$ molecules in their neighborhood which is enough for branches and
film to grow with the same rate.

The dependence of $r_A$ on $N_0$ is not explicit in the model, but it exists. As
the number of branches per unit area increases, the aggregation of the
diffusing additive to them is facilitated and, consequently, their growth rate
$r_A$ increases. From the dependence of the saturation value of $H-h$ on $a$,
we expect that it decreases as $N_0$ increases: the steady state regime is more
rapidly attained ($1/a$ decreases), with a smaller exposed height of the
branches. On the other hand, the dependence of $H-h$ on $r_A$ suggests the
opposite behavior, i. e. that it increases with $N_0$. The interplay of these
effects will determine the overall dependence of $H-h$ on $N_0$. Anyway,
this analysis shows the nontrivial consequences of the nucleation processes
(which determine $N_0$) on the long time behavior of the system.

Now let us consider the effect of the additive concentration in solution,
$\rho_A$, assuming that its diffusion coefficient is constant (constant
temperature conditions). Since $r_0$ only represents the reactions occurring
near the film surface, it is expected to be constant as $\rho_A$ increases. On
the other hand, the growth rate of the branches, $r_A$, is some monotonically
increasing function of $\rho_A$. For low concentrations,
$r_A$ is small, consequently the branches will be covered by $ZnO$.
Quantitatively, this is the case where $r_A\left( \rho_A\right) <r_0$, which
implies $\rho_A<\rho_c$, where $\rho_c$ is a critical concentration of additive
such that
$r_A\left( \rho_c\right) =r_0$. For higher concentrations, we have
$r_A>r_0$ and the above solution of the rate equation model is valid.

The dependence of the growth rate on the additive concentration, with the
catalytic effect only for $r_A>r_0$, is illustrated in Fig. 3. There, we assume
that $r_A$ linearly increases with
$\rho_A$ when the concentration is not too large (first order reaction). This is
the regime where our model is expected to work. For large concentrations, Fig.
3 shows a saturation of growth rate, which is a consequence of
the limitations to oxigen diffusion towards the surface. Indeed, in this
regime, the catalytic effect of eosin to reduce oxygen does not result in
an increased production of hydroxide ions anymore. The hydroxide precursor
current towards the electrode becomes constant, which also occurs with the
growth rate of ZnO, which is directly linked to the availability of hydroxyde
ions.

The behavior observed in the intermediate EY concentration range is in
qualitative agreement with the findings
of Ref. \protect\cite{pauporte4} for $ZnO$/EY electrodeposition with
hydrogen peroxide as hydroxide precursor: no significant effect in the growth
rate is found for EY concentrations below $2\mu M$, but an increase
is observed for EY concentrations above that value. On the other hand, no
well-defined plateau in the growth rate is observed in electrodeposition with
molecular oxygen \cite{pauporte8}, but an increase in the growth rate even for
small additive concentrations. This is probably due to nucleation features that
are not properly represented in the present simplified model.

The main parameters of the rate equation model are the growth rates $r_A$ and
$r_0$, which depend on several parameters, such as
concentrations of hydroxide precursors and additives in solution, zinc
ions concentration and the respective diffusion coefficients. However, the model
leads to two important conclusions. First,
in order that the catalytic effect is observed, it is necessary that $r_A>r_0$,
consequently the concentration and the diffusion coefficient of the additive in
solution must be combined in a suitable way to fullfill this condition.
Secondly, if the additive branches are able to grow above the film surface,
then the film growth rate is the same of those branches in the stationary state
($t\gg 1/a$ in Eq. \ref{ht}), independently of the rate of pure $ZnO$ films
growth. These conclusions will guide the simulation work on the microscopic
model.

\section{Microscopic model}

Now we present a more complete model for the electrodeposition of $ZnO$ with
organic additives, in which the physical properties of different species in
solution and in the deposit are taken into account. This is particularly
important to understand the role of diffusion coefficients and to predict the
structure of the deposit, as well as to
perform quantitative comparisons with experimental data. At the latter point,
we will basically refer to $ZnO$/EY films data available, thus the
presentation of the model will be inspired by this growth process.

We will consider a lattice model with two species executing random walks in
solution and two species forming a deposit grown above a flat electrode. All
lengths will be given in lattice units. Due to the relative complexity of this
model, it will be solved by simulation. A two-dimensional version will be
discussed, which we expect to capture the main features of the real process,
at the same time of being amenable for computation.

The species in the solution, O and E, represent the hydroxide precursor ($O_2$
or $H_2O_2$) and eosin Y, respectively.
Their different sizes, as shown in Fig. 4a, qualitatively account for the
different masses and radius. Their auto-diffusion coefficients are $D_O$
and $D_E$, i. e. each O (E) in solution executes $D_O$ ($D_E$) random movements
of unit size per unit time. These rates set the time scale of our model. 

The deposit will be formed by two aggregated species, named ZA and EA,
representing respectively the $ZnO$ and the eosin Y molecules in the
film, as shown in Fig. 4b. For simplicity, their sizes are the same as those of
the species in solution.

Although the zinc and hydroxide ions play an essential role in
the reactions leading to the formation of $ZnO$ (Eqs. \ref{reacoxigen},
\ref{reacperoxide} and \ref{reaczno}), they will not be explicitely represented
in our model. Instead, their effects will be described by probabilities of
removing O particles from the solution and producing new aggregated particles,
thus avoiding the complications of the explicit representation of the above
mentioned reactions. Following this reasoning, the growth process is
represented by the simplified reactions $O\to ZA$ and $E\to EA$, which can take
place with given probabilities  and under conditions that are related to the
neighborhood of the particles in solution.

The growth process begins with a flat electrode (a line of length $L$) and a
solution with concentrations $\rho_O$ and $\rho_E$ of O and E particles,
respectively.

The aggregation of an E particle (reaction $E\to EA$) occurs when that particle
has a nearest neighbor of a previously aggregated EA (Fig. 5a) or when it has a
nearest neighbor site belonging to the electrode (the initial substrate). The
new EA particle permanently sticks at the position where it is
formed. The contact of E particles with ZA sites never generates EA particles,
which is a key hypothesis for explaining the formation of continuous fibers of
eosin Y in the films.

The reaction $O\to ZA$ occurs upon contact of a particle O
with ZA or EA. When the O particle has a neighbor site ZA or a neighbor site
belonging to the
electrode (initial substrate), the reaction occurs with probability $p$,
otherwise the O particle
remains in the solution. When the O particle has a neighboring EA, that reaction
occurs with probability $1$. These rules are illustrated in Fig. 5b.

Contrary to the EA particles, ZA particles are allowed to precipitate and
diffuse on the film surface after their formation. In order to represent the
main features expected for this
process and save computational time, a simple diffusion mechanism will be
mimicked immediately after the ZA particle is created: it precipitates to the
topmost empty site of its current column, searches for the
point with the highest number of nearest neighbors ZA whithin a certain radius
$R$ (defined below), and permanently sticks at that point. These processes are
illustrated in Fig. 5c.

The above stochastic rules are motivated by some experimental features of the
$ZnO$/EY films growth. The rules for formation of EA particles represent
the initial nucleation of EY at the electrode and the subsequent formation
of EY clusters upon reduction of the molecules that reach the film
surface. Since EA is not formed upon contact of E with ZA, there
will be no isolated EA particle in the ZA matrix, which accounts for the fact
that almost all EY can be removed from the hybrid films, leaving them with a
porous structure \cite{yoshida5,pauporte7}. As regards the formation of ZA
particles, the higher
probability of occurrence after contact of O with EA follows from the
assumption that oxygen precursors reduction is enhanced near the eosin
clusters, which facilitates the formation of $ZnO$ in those regions. This
accounts for the catalytic effect of eosin Y, also observed experimentally.

For the diffusion of a ZA particle on the ZnO film surface after its formation,
we consider a surface diffusion
coefficient $D_A$, which gives a radius for searching the final
aggregation position as $R=\sqrt{D_A\tau}$, where $\tau$ is the
current average time for one layer deposition. This time is updated during the
simulation, while $D_A$ is kept fixed. The condition to choose the final
aggregation point follows the same ideas of the Wolf-Villain model for thin film
deposition \cite{wv}, and are
reasonable for formation of a crystalline structure due to the trend of
increasing the binding energy.

The height of the solution above the film surface is kept constant during the
simulation of the growth process. New particles (O or E) are left at random
positions of the top layer of the solution immediately after an aggregation
event (rections $E\to EA$, $O\to ZA$), in order to maintain constant
concentrations. This height was $300$ lattice units in our simulations. The
diffusion layer near the film surface, in which concentrations of O and E were
reduced, was of order of $100$ lattice
units or less. These conditions are reasonable to represent experiments in which
a rotating disk electrode was used, thus avoiding the growth
of the diffusion layer and the subsequent decrease of the growth rate.

We recall that the rules for aggregation of E particles are the same of the
original diffusion-limited aggregation (DLA) model of Witten and Sander
\cite{witten}, in which growth took place from a single seed. Extensions to the
case of an initial flat surface were also considered by several authors
\cite{meakin,racz,jullien,burlatsky}. To be more precise, due to the collective
diffusion mechanism of E particles in solution, the growth kinetics of E is
equivalent to that of the multiparticle biased DLA (MBDLA) in the condition of
zero bias \cite{sanchez,castro1,castro2,schwarzacher}. Indeed, MBDLA was
proposed as a model
for electrodeposition of $CoP$ alloys \cite{castro1,riveiro}. On the other
hand, the aggregation rules of ZA particles, which include surface relaxation,
suggest Edwards-Wilkinson growth in that case \cite{ew,barabasi}. However, in
the cooperative growth
model, a nontrivial interplay of both processes is observed, as shown below.

\section{Results for the microscopic model}

In all simulations presented here, we considered $D_O=1$, and in most cases the
lattice lenghts are $L=512$ and the diffusion coefficient of aggregated ZA is
$D_A=0.005$. Some data for $L=2048$ were also
collected in order to search for possible finite-size effects, mainly when
the thickest deposits were grown.
In experiments, the maximum ratio of concentrations of eosin and hydroxide
precursor in
solution is in the range ${10}^{-3}-{10}^{-2}$, thus in this model we worked
with ratios between $0$ (no eosin) and ${10}^{-2}$.

The results of the rate equation model of Sec. II suggest that the growth rate
$r_A$ of the additive (eosin) is proportional to $\rho_E D_E$, while
the growth rate of the pure $ZnO$ deposit, $r_0$, is proportional to $p \rho_O
D_O$. In the following, the model parameters will be tuned 
in the light of these relations and the conclusions of the model
of Sec. II.

First the model was simulated considering $D_E=1/8$ (thus $D_E/D_O\sim 0.1$), in
order to account for the
much larger mass of eosin when compared to the hydroxide precursors.
$D_A=0.05$ was considered in all
simulations, some of them with $\rho_O=2\times {10}^{-2}$,
$\rho_E=2\times {10}^{-4}$, and others with $\rho_O=5\times {10}^{-3}$,
$\rho_E=5\times {10}^{-5}$. Several different values of $p$ were considered,
with $p\geq {10}^{-5}$.

In Fig. 6 we show a region of a deposit obtained with $p= {10}^{-4}$,
$\rho_Z=5\times {10}^{-3}$ and $\rho_E=5\times {10}^{-5}$, at
$t=3\times {10}^5$. Some EA
particles are present at the initial electrode surface, but they are rapidly
covered by the ZA layer. Subsequently, there is no possibility for uptaking
other eosin molecules in the film, then the
final growth rate is determined by ZA aggregation. From the results obtained
with such small diffusion coefficient of E in solution, we conclude that this
is not the suitable condition to observe the catalytic effect of the additive.

Subsequently, we assumed that E particles had a significantly large diffusion
coefficient, $D_E=1$, which is equal to the coefficient of Z.
In Figs. 7a, 7b and 7c, we show the time evolution of a region of a deposit
grown with $\rho_O=2\times {10}^{-2}$,
$\rho_E=2\times {10}^{-4}$ and $p={10}^{-3}$. Although many EA particles attach
to the electrode in the beginning of the process, they frequently
become covered by ZA particles after some time, even with a small
probability of O-ZA aggregation. The growth process is improved only at isolated
regions where clusters of EA particles succeed to grow, as shown in
Figs. 7a-7c.

When compared to the case without E particles, the average height
of the deposit increases more than twice near the clusters of Figs. 7a-c, for
the same time of growth. However, the large distance between the surviving
branches leads to a very large surface roughness at long lengthscales. 
Moreover, this roughness rapidly increases in time because the mounds growing
with large rates are separated by long valleys which slowly grow. These
features are not observed in experiments, for which we refer to Fig. 1 as a
typical example. We also tested smaller concentrations of $\rho_O$, but keeping
the relative concentration between E and O fixed (${10}^{-2}$), obtaining
similar results (growth of EA branches may be even more difficult for the same
$p$). We conclude that larger values of $D_E$ must be tested in order to find
the suitable conditions to represent the experimental results.

Following this reasoning, we also considered the case $D_E=2$, and obtained
results that qualitatively agree with experimental findings. In this case,
small
concentrations of E particles are able to improve the growth process, with the
creation of a large number of nucleation centers in the eletrode, the
formation of a large number of branches and a competition between them as the
deposit grows, as well as a much smaller surface roughness.
The formation of an eosin interconnected network inside the ZnO matrix is clear,
which also agrees with the main features of the experimental nanostructures.

Fig. 8 illustrates the beginning of the growth of a region of a deposit with
$\rho_O=5\times {10}^{-3}$,
$\rho_E=5\times {10}^{-5}$, $D_A=0.05$ and $p=0.25$.
The average height for $t=3\times {10}^6$ (Fig. 8c) is nearly $3$ times larger
than the height of the film grown without E particles during the same time, and
the eosin loadings in the deposits  (ratio between number of EA particles and
ZA particles) of Figs. 8a-8c increase from $0.64\%$ to
$0.85\%$. Simulation at longer times show sthat these quantities attain
different steady state values, so that the features of Fig. 8 may be
interpreted as typical of the initial growth process.

One interesting point revealed in Fig. 8 is the competition between the growing
EA clusters. The central cluster of Figs. 8a and 8b grew slower than the
neighboring ones and, consequently, was covered by particles ZA in Fig. 8c,
while the larger surviving clusters created more branches.

In Fig. 9 we show a region of a deposit grown until $t=2\times {10}^7$
with the same parameters of those in Fig. 8, except that $D_A=0.01$. The
surface roughness is larger  in this case due to the inhomogeneous growth of EA
branches and smaller diffusion coefficient of aggregated particles. However,
other quantities are not much different from those obtained with $D_A=0.05$,
such as the growth rate at steady state conditions, which is nearly $3.3$ times
larger than the rate of film grown without E particles. It clearly shows the
catalytic effect of the eosin clusters. Another important quantity is the
EA loading in the film, which is nearly $2\%$. This is a high value
compared to the relative concentration of $1\%$ in solution.

The above results show that a realistic description of the growth of
$ZnO$/EY films must take into account a remarkably large diffusion
coefficient for eosin in solution. A quantitative comparison with the ratio
$D_O/D_E$ of the model is not reliable because the model contains a small
number of adjustable parameters, while
the efficiency of other processes such as $ZnO$ precipitation are not taken into
account. However, it is noticeable that an unexpectedly large diffusion
coefficient of EY is also obtained experimentally:
$3.4\times {10}^{-5} cm^2 s^{-1}$ and $1.4\times {10}^{-5} cm^2 s^{-1}$ for
$O_2$ and EY, respectively, in the classical deposition condition of the films,
that is at $70{}^oC$ and in chloride medium \cite{pauporte4}. Consequently, the
experimental ratio between diffusion coefficients is near $2.4$, a much smaller
value than the ratio between the masses of EY and $O_2$, which is $20$. 

The order of magnitude of the growth rate increase in the films of Figs. 8 and
9, when compared to the films
without EA particles, is also consistent with experiments: a factor $3$ was
obtained in Ref. \protect\cite{pauporte4}, with hydrogen peroxide as the oxygen
precursor, when $\rho_E/\rho_O = {10}^{-3}$, while a factor
$5.5$ was obtained in Ref. \protect\cite{pauporte8}, with molecular oxygen
precursor, when $\rho_E/\rho_O = {10}^{-2}$. Moreover, the high eosin loading
of $2.1\%$ was observed in the latter experiments \cite{pauporte8}, which also
agrees with the predictions of our model.

We also recall that the rules of the model were proposed so that all EA
branches are connected to the electrode - see the images in Figs. 7, 8 and 9.
At long times, it is also
observed that most EA particles belong to some of the branches which are
exposed above the film surface - see Figs. 8 and 9. Accordingly, a small
fraction of EA particles is hidden in the deposit. This also agrees with the
experimental finding that eosin can be almost completely removed by a soft
chemical treatment of the hybrid films \cite{yoshida5,pauporte7}.

Finally, it is interesting to point out that the configurations of EA
branches are similar to those encountered in a model of diffusion-controlled
deposition by Burlatsky et al \cite{burlatsky}, whose growth mechanisms are the
same adopted for the E particles in our model. However, while the screening
process of the larger clusters was the only reason for small clusters to stop
growing in their model, here this effect is enhanced by the deposition
of ZA particles, which may eventually cover an EA cluster and suppress its
growth (see e. g. Fig. 8c). It is also important to notice the similarity
between the EA clusters and the electrodeposited copper aggregates of Refs.
\protect\cite{leger1,leger2}, which clearly show such screening effects.

From the point of view of competitive growth models, the results of the present
paper also show nontrivial features. It is usually observed that one of the
competing dynamics is dominant at long times and large lengthscales, such as in
the widely studied EW-KPZ crossover \cite{GGG,NT,AF,chamereis} or in random to
correlated growth \cite{rdcor}. However, here it was shown that a less trivial
association of these different dynamics to different chemical species leads to
a cooperative behavior where features of both dynamics are to some extent
preserved. In the present model, the competing dynamics are those of MBDLA
without bias for the EA particles, and of the EW model for the ZA particles, as
discussed in Sec. IV.

\section{Conclusion}

We proposed a rate equation model and a microscopic statistical model to
represent some features of $ZnO$ electrodeposition with organic additives,
mainly focusing the application to electrodeposition with eosin Y.

The rate equation model assumes the formation of branches of the additive which
are extended above the film surface, and is useful to understand the basic
features of the growth process. Under conditions that the branches grow faster
than the pure $ZnO$ deposit, it was shown that both structures grow with the
growth rate of the former. 

With the statistical model, we were able to reproduce several qualitative
features of the
electrodeposition of $ZnO$/EY films by assuming that it may be viewed as a
diffusion-limited aggregation process with an interplay between different
chemical species in solution and a catalytic effect of the eosin clusters for
the formation of $ZnO$. Among the predictions of the model which are in
qualitative agreement with experimental findings, it is
important to mention: 1) the requirement of particularly large diffusion
coefficients of EY in solution, so that it is able to improve the growth
process with small concentrations; 2) the formation of
branches of EY connected to the surface
and filling the film structure without significant large scale inhomogeneities;
3) the increase in relative growth rates and the high EY loadings.
Consequently, we believe that this model incorporated most of the basic features
of the $ZnO$ electrodeposition with EY, and may be extended in order to
provide quantitatively good results or to represent related growth processes.

\vskip 1cm

{\bf Acknowledgements}

The authors thank Dr. Aur\'elie Goux for the preparation of the film shown in
Fig.1. FDAAR thanks Laboratoire de Electrochimie et Chimie Analytique where part
of this work was done, for the
hospitality, and acknowledges support by CNPq (Brazil) and CNRS (France).
DL and TP acknowledge the fruitful collaboration with Prof. Tsukasa Yoshida
devoted to the electrodeposition of ZnO-Eosin films.


\vfill\eject

\begin{figure}[!h]
\centering
\includegraphics[clip,
width=\textwidth,
height=0.85\textheight,
angle=0]{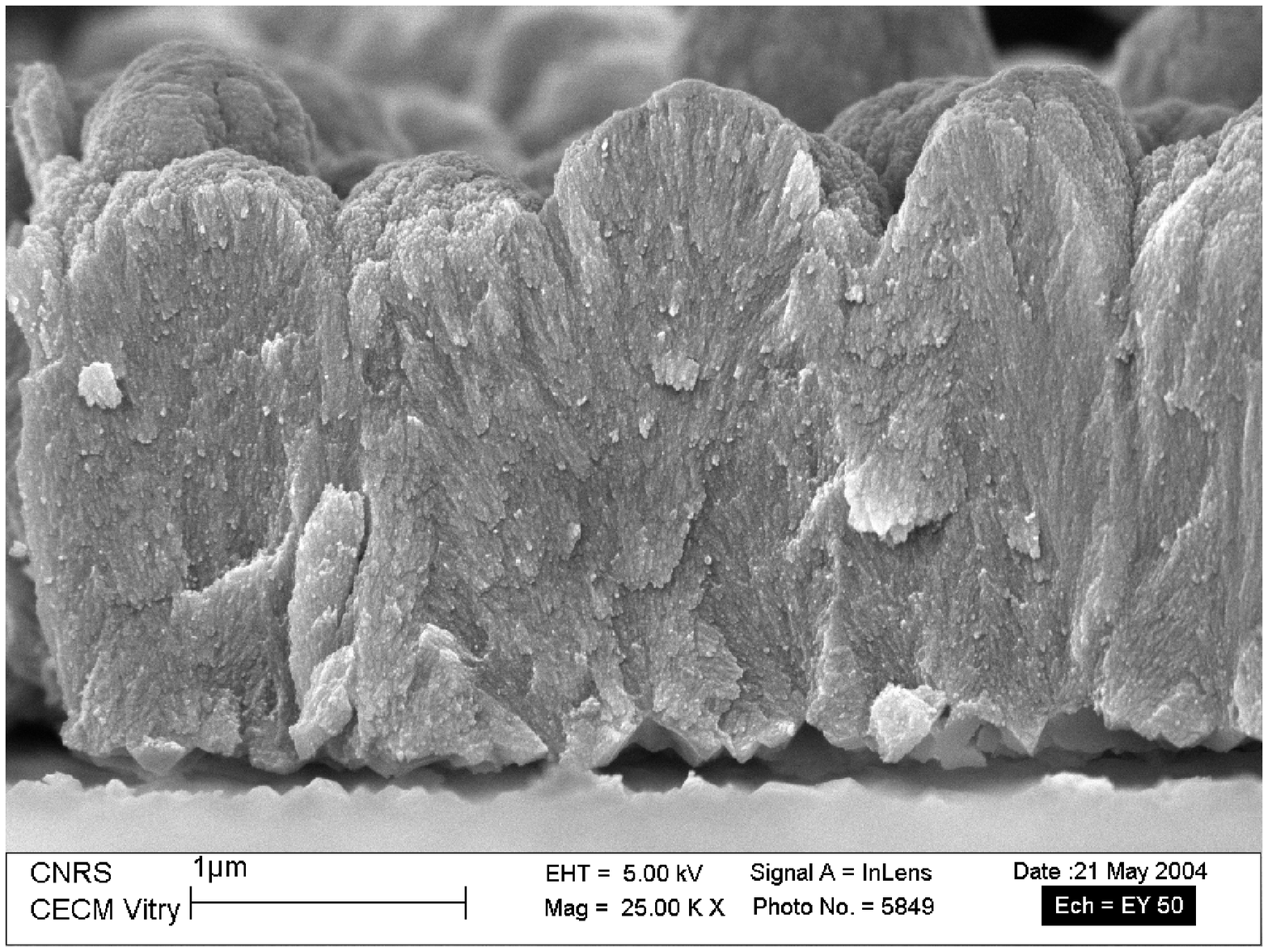}
\caption{\label{fig1} FESEM cross-sectional view of a $ZnO$/EY thin film
prepared by electrodeposition.}
\end{figure}

\begin{figure}[!h]
\includegraphics[clip,width=0.80\textwidth,
height=0.25\textheight,angle=0]{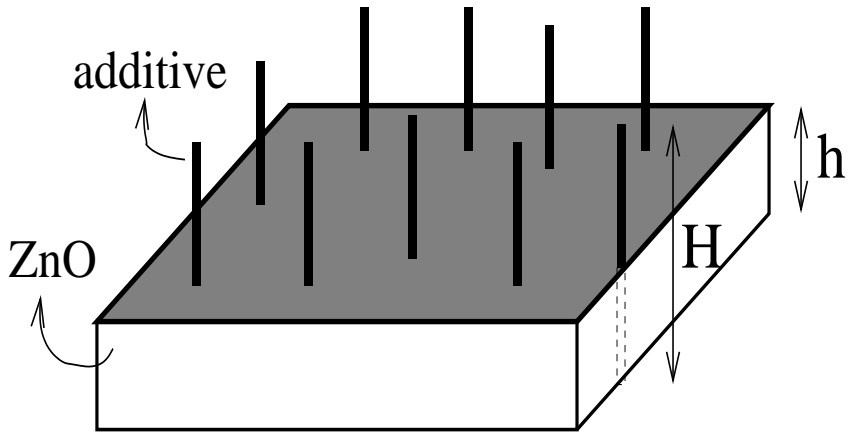}
\caption{\label{fig2} Scheme for the rate equation model, with the $ZnO$ film of
height $h$
growing from the electrode below it, and straight branches of an additive of
height $H$.}
\end{figure}

\vskip 5cm

\begin{figure}
\includegraphics[clip,width=0.70\textwidth, 
height=0.35\textheight,angle=0]{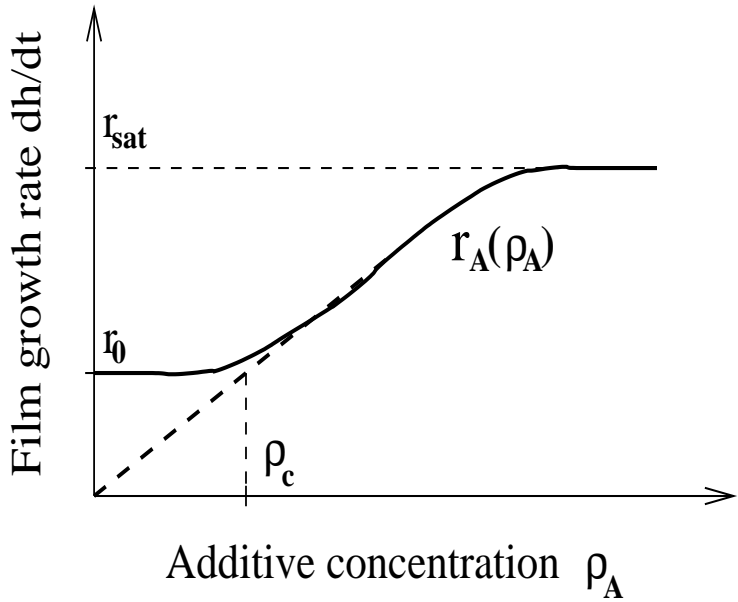}
\caption{\label{fig3} The solid curve shows the $ZnO$ film growth rate as a
function of the additive concentration in the rate equation model. The dashed
curve is an extension of the solid curve for $\rho_A <\rho_c$ [curve
$r_A\left( \rho_A \right)$]. For higher concentrations of the additive, the
diffusion of the oxygen precursor becomes limiting, which corresponds to the
plateau in the growth rate.}
\end{figure}

\begin{figure}[!h]
\includegraphics[clip,width=0.80\textwidth, 
height=0.42\textheight,angle=0]{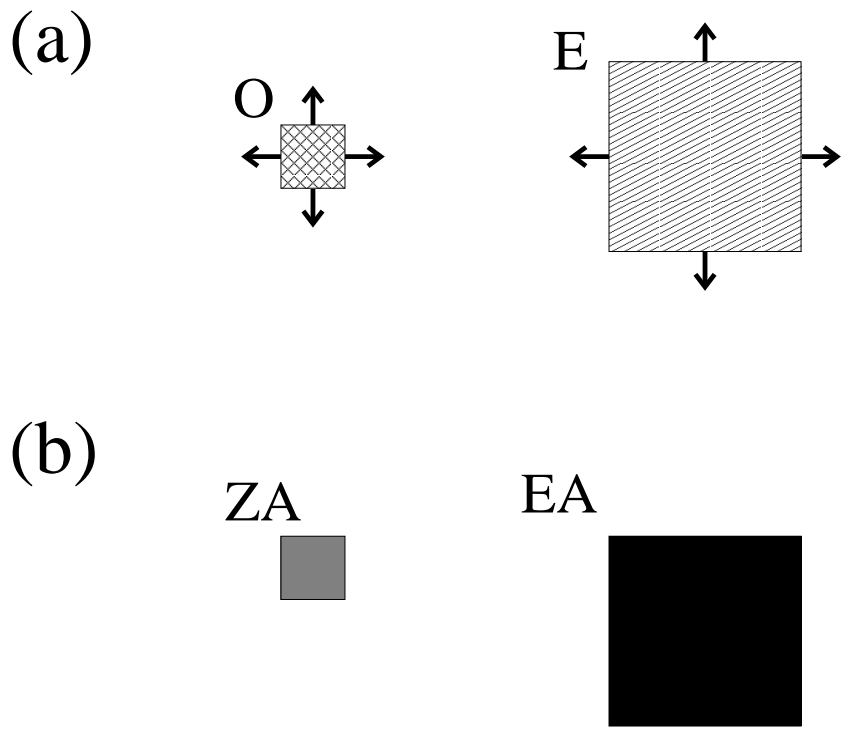}
\caption{\label{fig4} (a) Species in the solution of the statistical model are
represented by squares
of sizes equal to $1$ lattice unit (O) and $3$ lattice units (E). (b)
Aggregated species (ZA and EA) are represented by squares of the same sizes of
the solution.}
\end{figure}

\begin{figure}[!h]
\includegraphics[clip,width=1.0\textwidth, 
height=0.45\textheight,angle=0]{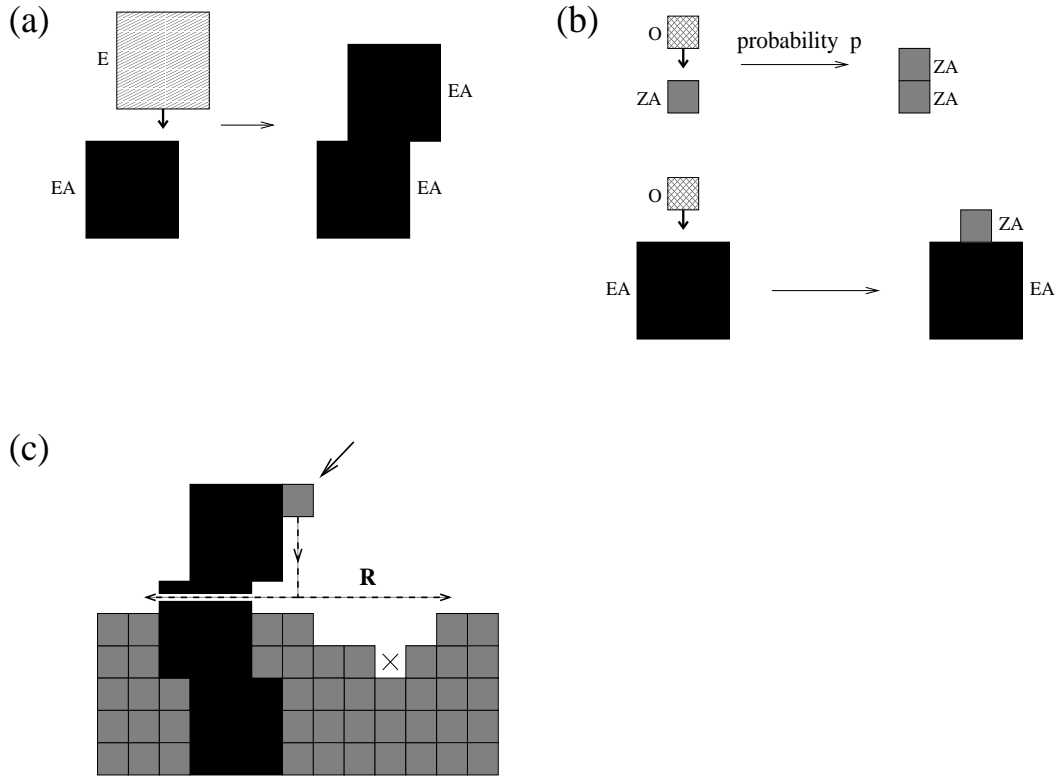}
\caption{\label{fig5} (a) Rule for aggregation of E particles ($E\to EA$), which
is possible only upon contact with an EA particle.
(b) Probabilistic rules for aggregation of O particles ($O\to ZA$). The
aggregation occur with probability $1$ in contact with EA.
(c) Precipitation and diffusion of a ZA particle (indicated by an arrow)
immediately after its creation.
$R$ is the radius to search for the final aggregation point. $\times$ indicates
the point to be chosen in this case, which presents the largest number of
neighbors ZA.
}
\end{figure}

\begin{figure}[!h]
\includegraphics[clip,width=\textwidth,
height=0.68\textheight,angle=0]{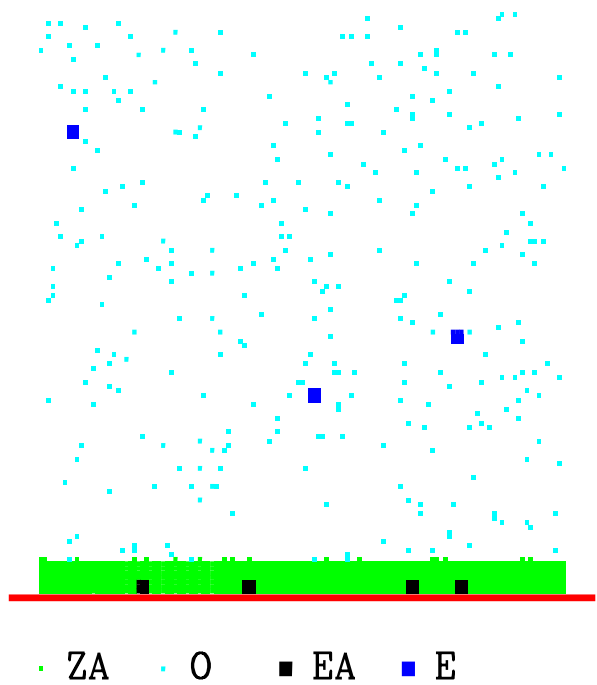}
\caption{\label{fig6} Section of lateral size $128$ of a deposit grown with
$D_O=1$, $D_E=1/8$, $\rho_O=0.005$, $\rho_E=0.00005$, $D_A=0.05$ and
$p={10}^{-4}$ at $t=3\times {10}^5$, with the solution above it. The horizontal
line at the bottom represents the electrode.}
\end{figure}

\begin{figure}[!h]
\includegraphics[clip,width=\textwidth,
height=0.68\textheight,angle=0]{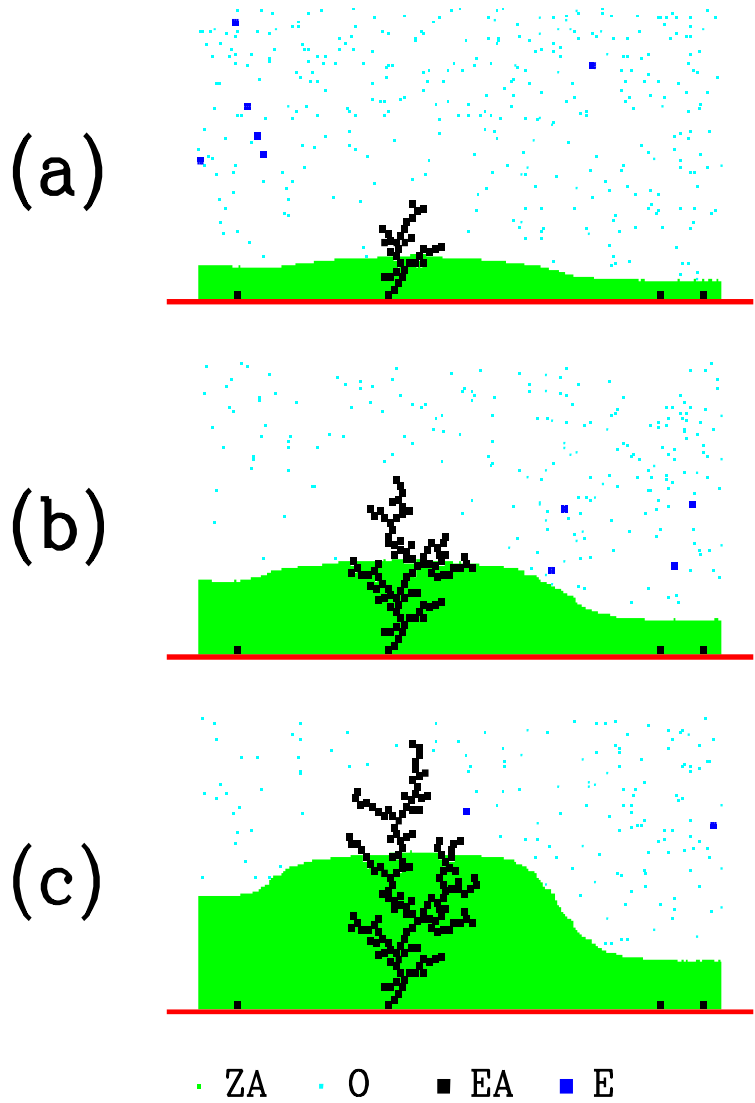}
\caption{\label{fig7} Initial evolution of a section of lateral size $256$ of a
deposit grown with $D_O=1$, $D_E=1$, $\rho_O=0.02$, $\rho_E=0.0002$, $D_A=0.05$
and $p=0.001$. Growth times are:
(a) $t=5\times {10}^5$, (b) $t={10}^6$ and (c) $t=1.5\times {10}^6$. The
horizontal
line at the bottom represents the electrode.}
\end{figure}

\begin{figure}[!h]
\includegraphics[clip,width=\textwidth,
height=0.68\textheight,angle=0]{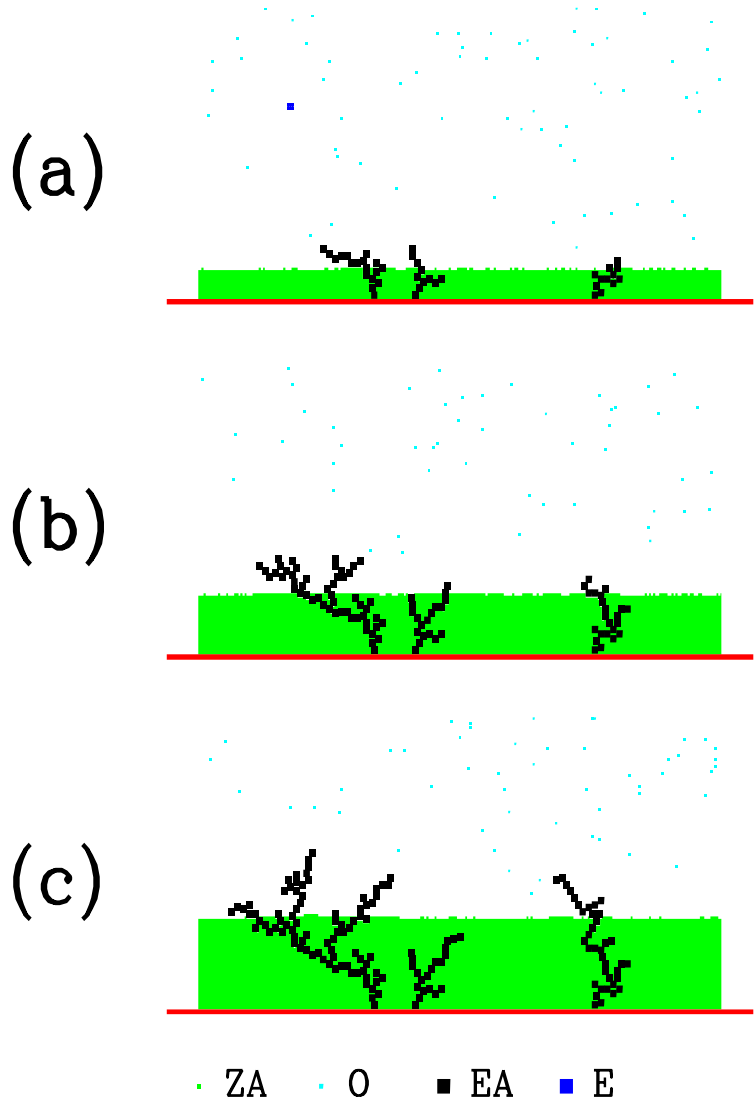}
\caption{\label{fig8} Initial evolution of a section of lateral size $256$ of a
deposit grown with $D_O=1$, $D_E=2$, $\rho_O=0.005$, $\rho_E=0.00005$,
$D_A=0.05$ and $p=0.25$. Growth times are:
(a) $t={10}^6$, (b) $t=2\times {10}^6$ and (c) $t=3\times {10}^6$. The
horizontal
line at the bottom represents the electrode.}
\end{figure}

\begin{figure}[!h]
\includegraphics[clip,width=\textwidth,
height=0.68\textheight,angle=0]{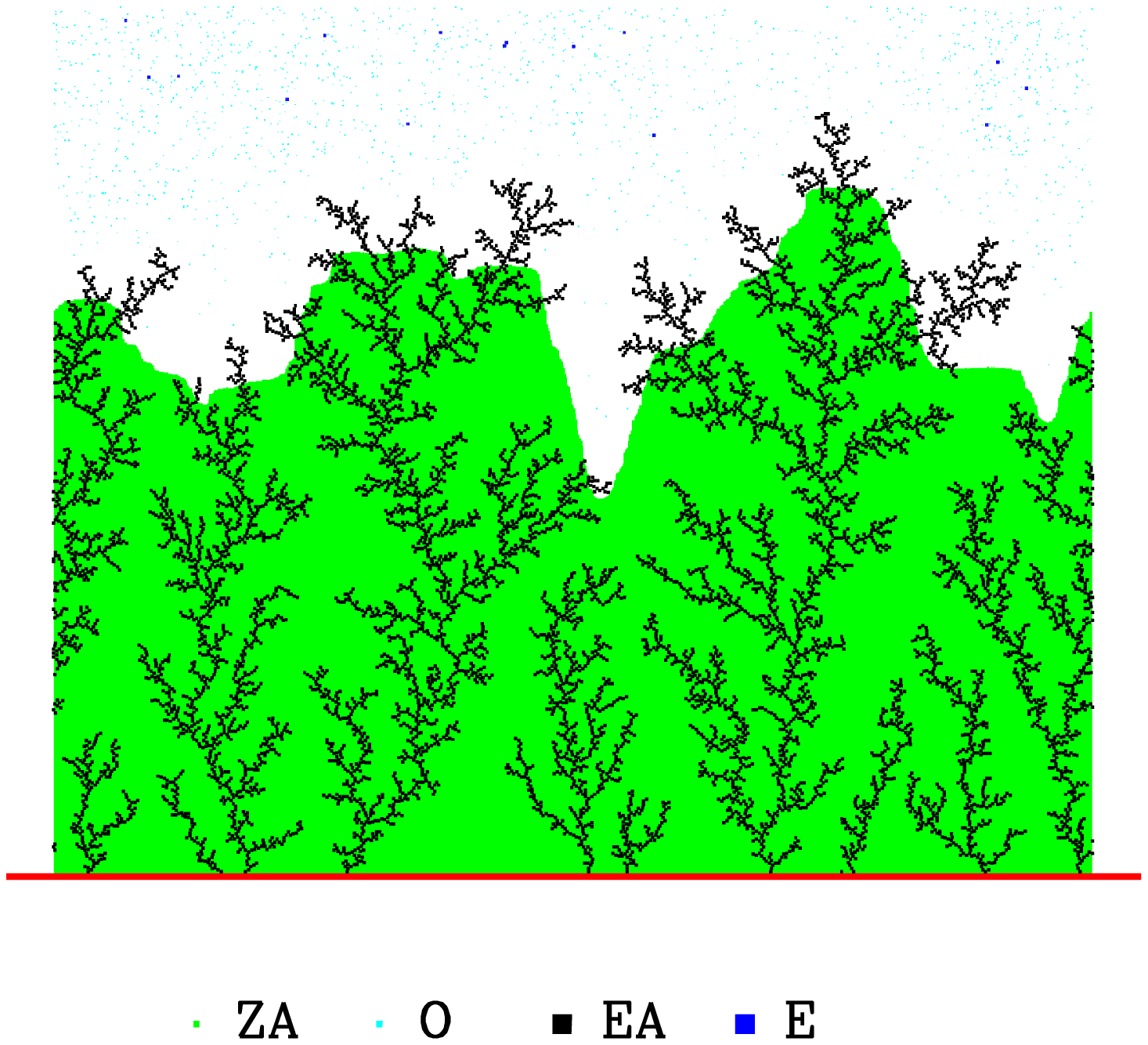}
\caption{\label{fig9} Section of lateral size $1024$ of a
deposit grown with $D_O=1$, $D_E=2$, $\rho_O=0.005$, $\rho_E=0.00005$,
$D_A=0.01$ and $p=0.25$, at $t=2\times {10}^7$. The
horizontal
line at the bottom represents the electrode.}
\end{figure}


\begin{thebibliography}{99}

\bibitem{revlincot}
D. Lincot, Thin Solid Films {\bf 487} (2005) 40.

\bibitem{emanetoglu}
N. W. Emanetoglu, C. Gorla, Y. Liu, S. Liang, and Y. Lu, Mater. Sci. Semicond.
Process {\bf 2} (1999) 247.

\bibitem{konenkamp}
R. Konenkamp, R.C. Word, and M. Godinez, Nanoletters {\bf 5} (2005) 2005.

\bibitem{huang}
M.H. Huang, S. Mao, H. Feick, H. Yan, Y. Wu, H. Kind, E. Weber, R. Russo, and P.
Yang, Science {\bf 292} (2001) 1897.

\bibitem{lin}
Y. Lin, Z. Zhang, Z. Tang, F. Yuan, and J. Li, Adv. Mater. Opt. Electron. {\bf
9}
(1999) 205.

\bibitem{golego}
N. Golego, S.A. Studenikin, and M. Cocivera, J. Electrochem. Soc. {\bf 147}
(2000)
1592.

\bibitem{keis}
K. Keis, C. Bauer, G. Boschloo, A. Hagfelt, K. Westermark, H. Rensmo, and H.
Siegbahn, J. Photochem. Photobiology A {\bf 148} (2002) 57.

\bibitem{vaysieres}
L. Vaysieres, Adv. Mater. {\bf 15} (2003) 464.

\bibitem{tian}
Z.R. Tian, J.A. Voigt, J. Liu, B. Mckenzie, M. J. Macdermott, M.A. Rodriguez,
H. Konishi, and H. Xu, Nature Mater. {\bf 2} (2003) 821.

\bibitem{izaki}
M. Izaki and T. Omi, Appl. Phys. Lett. {\bf 68} (1996) 2439.

\bibitem{peulon1}
S. Peulon and D. Lincot, Adv. Mater. {\bf 8} (1996) 166

\bibitem{peulon2}
S. Peulon and D. Lincot, J. Electrochem. Soc. {\bf 145} (1998) 864.

\bibitem{pauporte1}                         
T. Pauport\'e and D. Lincot, Appl. Phys. Lett. {\bf 75} (1999) 3817.

\bibitem{pauporte2}
T. Pauport\'e and D. Lincot, J. Electrochem. Soc. {\bf 148} (2001)  C310.

\bibitem{pauporte3}
T. Pauport\'e and D. Lincot, J. Electroanal. Chem. {\bf 517} (2001) 54.

\bibitem{choi}
K.S. Choi, H.C. Lichtenegger, and G.D. Stucky, J. Am. Chem. Soc. {\bf 124}
(2002)
12402.

\bibitem{michaelis}
E. Michaelis, D. W\"ohrle, J. Rathousky, and M. Wark, Thin Solid Films {\bf 497}
(2006) 163.

\bibitem{yoshida1}
T. Yoshida, K. Terada, D. Schlettwein, T. Oekermann, T. Sugiura, and H. Minoura,
Adv. Mater. {\bf 12} (2000) 1214.

\bibitem{pauporte4}
T. Pauport\'e, T. Yoshida, A. Goux, and D. Lincot, J. Electroanal. Chem. {\bf
534} (2002) 55.

\bibitem{yoshida2}
T. Yoshida, T. Pauport\'e, D. Lincot, T. Oekermann, and H. Minoura, J.
Electrochem.
Soc. {\bf 150} (2003) C608.

\bibitem{pauporte5}
T. Pauport\'e, T. Yoshida, R. Cort\`es, M. Froment , and D. Lincot, J. Phys.
Chem.
B, {\bf 107} (2003) 10077.

\bibitem{okabe}
K. Okabe, T. Yoshida, T. Sugiura, and H. Minoura, Trans. Mater. Research. Soc.
Jap.,
{\bf 26} (2001) 523.

\bibitem{yoshida3}
T. Yoshida, K. Miyamoto, N. Hibi, T. Sugiura, H. Minoura, D. Schlettwein, T.
Oekermann, G. Schneider, and D. W\"ohrle, Chem. Lett. {\bf 27} (1998) 599.

\bibitem{yoshida4}
T. Yoshida, M. Tochimoto, D. Schlettwein, D. W\"ohrle, T. Sugiura, and H.
Minoura,
Chem. Mater. {\bf 11} (1999) 2657.

\bibitem{schlettwein}
D. Schlettwein, T. Oekermann, T. Yoshida, M. Tochimoto, and H. Minoura, J.
Electroanal. Chem. {\bf 481} (2000) 42.

\bibitem{pauporte6}
T. Pauport\'e, F. Bedioui, and D. Lincot, J. Mat. Chem. {\bf 15} (2005) 1552

\bibitem{karuppuchamy}
S. Karuppuchamy, T. Yoshida, T. Sugiura, and H. Minoura, Thin Solid Films {\bf
397}
(2001) 63.

\bibitem{goux}
A. Goux, T. Pauport\'e, J. Chivot, and D. Lincot, Electrochim. Acta {\bf 50}
(2005)
2239.

\bibitem{yoshida5}
T. Yoshida, M. Iwaya, H. Ando, T. Oekermann, K. Nonomura, D. Schlettwein, D.
W\"ohrle, and H. Minoura, Chem. Comm. (2004) 400.

\bibitem{pauporte7}
T. Pauport\'e, T. Yoshida, D. Komatsu, and H. Minoura, Electrochem. Solid State
Lett. {\bf 9} (2006) H16.

\bibitem{lincot1}
D. Lincot, T. Pauport\'e, A. Goux, V. Lair, and T. Yoshida,
Meet. Abstr. - Electrochem. Soc. {\bf 501}, 460 (2006).

\bibitem{pauporte8}
T. Pauport\'e, unpublished results.

\bibitem{wv}
D. Wolf and J. Villain, Europhys. Lett. {\bf 13} (1990) 389.

\bibitem{witten}
T. A. Witten and L. M. Sander, Phys. Rev. Lett. {\bf 47} (1981) 1400.
                                                         
\bibitem{meakin}
P. Meakin, Phys. Rev. A {\bf 27}, 2616 (1983); {\bf 30} (1984) 4207.

\bibitem{racz}
Z. Racz and T. Vicsek, Phys. Rev. Lett. {\bf 51} (1983) 2383.

\bibitem{jullien}
R. Jullien, M. Kolb, and R. Botet, J. Physique (Paris) {\bf 45} (1984) 395.

\bibitem{burlatsky}
S. F. Burlatsky, G. S. Oshanin, and M. M. Elyashevich, Phys. Lett. A {\bf 151},
(1990) 538.

\bibitem{sanchez}
A. S\'anchez, M. J. Bernal, and J. M. Riveiro, Phys. Rev. E {\bf 50} (1994)
R2427 .

\bibitem{castro1}
M. Castro, R. Cuerno, A. S\'anchez and F. Dom\'{\i}nguez-Adame, Phys. Rev. E
{\bf 57} (1998) R2491.

\bibitem{castro2}
M. Castro, R. Cuerno, A. S\'anchez and F. Dom\'{\i}nguez-Adame, Phys. Rev. E
{\bf 62} (2002) 161.

\bibitem{schwarzacher}
W. Schwarzacher, J. Phys.: Condens. Matter {\bf 16} (2004) R859.

\bibitem{riveiro}
J. M. Riveiro and M. J. Bernal, J. Non-Cryst. Solids {\bf 160} (1993) 18.

\bibitem{ew}
S. F. Edwards and D. R. Wilkinson, Proc. R. Soc. London {\bf 381}  (1982) 17.

\bibitem{barabasi}
A. L. Barab\'asi and H. E. Stanley, {\it Fractal concepts in surface growth}
(Cambridge University Press, Cambribge, England, 1995).

\bibitem{leger1}
C. L\'eger, J. Elezgaray, and F. Argoul, Phys. Rev. Lett. {\bf 78} (1997) 5010.

\bibitem{leger2}
C. L\'eger, J. Elezgaray, and F. Argoul, Phys. Rev. Lett. {\bf 58} (1998) 7700.

\bibitem{GGG}
B. Grossmann, H. Guo, and M. Grant, Phys. Rev. A {\bf 43} (1991) 1727.

\bibitem{NT}
T. Nattermann and L.-H. Tang, Phys. Rev. A {\bf 45} (1992) 7156.

\bibitem{AF}
J. G. Amar and F. Family, Phys. Rev. A {\bf 45} (1992) R3373.

\bibitem{chamereis}
A. Chame and F. D. A. Aar\~ao Reis, Phys. Rev. E {\bf 66} (2002)
051104. 

\bibitem{rdcor}
F. D. A. Aar\~ao Reis, Phys. Rev. E {\bf 73} (2006) 021605.

\end{thebibliography}
\end{document}